\documentclass[%
superscriptaddress,
preprint,
 amsmath,amssymb,
 aps,
pra,
]{revtex4-1}

\usepackage{graphicx}% Include figure files
\usepackage{dcolumn}% Align table columns on decimal point
\usepackage{bm}% bold math
\usepackage{booktabs}
\usepackage{enumitem}
\usepackage{xcolor}
\begin{document}

%\preprint{APS/123-QED}

\title{Uncertainty Quantification for Misspecified Machine Learned Interatomic Potentials}

\author{Danny Perez}
 \email{danny\_perez@lanl.gov}
 \affiliation{Theoretical Division T-1, Los Alamos National Laboratory, Los Alamos, NM, 87544, USA}%Lines break automatically or can be forced with \\

 \author{Aparna P. A. Subramanyam}
 \affiliation{Theoretical Division T-1, Los Alamos National Laboratory, Los Alamos, NM, 87544, USA}%Lines break automatically or can be forced with \\

\author{Ivan Maliyov}%
\affiliation{Aix-Marseille Universit\'{e}, CNRS,
CINaM UMR 7325, Campus de Luminy, 13288 Marseille, France}

 \author{Thomas D Swinburne}%
 \email{thomas.swinburne@cnrs.fr}
\affiliation{Aix-Marseille Universit\'{e}, CNRS, CINaM UMR 7325, Campus de Luminy, 13288 Marseille, France}
%\affiliation{Department of Mechanical Engineering, University of Michigan, Ann Arbor, Michigan 48109, USA}
% (before I forget to do it in anticipation of grants etc....)

\date{\today}% It is always \today, today,

\begin{abstract}
    The use of high-dimensional regression techniques from machine learning
    has significantly improved the quantitative accuracy of interatomic 
    potentials. Atomic simulations can now plausibly target quantitative 
    predictions in a variety of settings, which has brought renewed interest 
    in robust means to quantify uncertainties on simulation results. 
    In many practical settings, encompassing both classical and a large class 
    of machine learning potentials, the dominant form of uncertainty is currently not due 
    to lack of training data but to misspecification, namely the inability of any one 
    choice of model parameters to exactly match all \textit{ab initio} training 
    data. However, Bayesian inference, the most common formal tool used to quantify uncertainty,  is known to ignore misspecification and 
    thus significantly underestimates parameter uncertainties. Here, we employ 
    a recent misspecification-aware regression technique to quantify 
    parameter uncertainties, which is then propagated to 
    a broad range of phase and defect properties in tungsten
    via brute force resampling or implicit differentiation. 
    The propagated misspecification uncertainties robustly envelope errors to direct \textit{ab initio} calculation of material properties outside of the training dataset, an essential requirement for any quantitative multi-scale modeling scheme. Finally, we demonstrate application to recent foundational machine learning interatomic potentials, accurately predicting and bounding errors in MACE-MPA-0 energy predictions across the diverse materials project database. Perspectives for the approach in multiscale simulation workflows are discussed.
    % DO WE TALK ABOUT INFO MAX TRAINING SET?
\end{abstract}

\maketitle

%\tableofcontents

\section{\label{sec:introduction}Introduction}
Atomic simulations such as molecular dynamics have long provided
detailed insight into the nanoscale dynamics of material behavior 
that would otherwise be extremely difficult to access in experiment.
For many years, these insights mostly took the form of mechanistic, qualitative 
information on the key nanoscale processes that dominate different material behavior. The focus on qualitative interpretation rather than quantitative 
predictions was a consequence of strong limitations in the accuracy of interatomic potentials, whose simple functional forms could not, nor 
could be expected to, deliver quantitative accuracy to some 
\textit{ab initio} reference calculations (typically obtained using Density Functional Theory, DFT), 
outside the small set of physical properties they were tailored to reproduce.

This outlook has gradually evolved with the introduction of machine learned interatomic potentials (MLIAP) \cite{behler2016perspective,deringer2019machine,mishin2021machine,goryaeva2021,lysogorskiy2021performant,mortazavi2023atomistic}, where expertly-crafted functional forms with a modest number of adjustable parameters were largely superseded by very flexible generic functional forms with a high number of adjustable parameters that can naturally capture very complex and non-intuitive chemical behavior. Early successes have led to a wave of optimism where the promise of "quantum accuracy at the cost of empirical potentials" was seen as being in reach (or indeed already achieved). It is now increasingly understood that the situation is more subtle: while MLIAPs can indeed deliver impressive accuracy, their flexibility and comparative lack of built-in physical constraints make the curation of the datasets used to train them a determining factor in their performance and robustness. For example, it was observed that MLIAPs can exhibit pathological behaviors, such as unstable dynamics, even when point-wise RMS and MAE errors are very low \cite{fu2023forces,Stocker_2022}. It was also observed that narrowly curated datasets can achieve very high local accuracy on configurations that are sufficiently similar to their training data, but exhibit large errors on more diverse datasets \cite{NPJ_Montes}, pointing to transferability challenges.

While the cost of generating high-accuracy electronic structure calculations 
has traditionally been a severe bottleneck limiting the complexity of IAPs, the 
exponential increase in available computing power \cite{alexander2020exascale} 
and improvements in the ability of electronic structure codes at exploiting 
modern hardware such as GPUs \cite{gavini2023roadmap}, has made the generation 
of extremely large training sets increasingly accessible. In conjunction with 
the fact that the community is increasingly seeking a more favorable compromise 
between accuracy and execution speed \cite{xie2023ultra}, the consideration of 
model errors in UQ for MLIAPs becomes increasingly urgent.

Because it is very difficult to {\em a priori} predict the types of configurations that will be encountered in simulations carried out by end-users, it is critical to provide robust, simple, and affordable methods to quantify the confidence in results produced by atomistic simulations based on MLIAPs. Uncertainty metrics can further be used to calibrate the IAPs themselves \cite{hegde2024bayesian} or to provide scoring functions that enable active learning approaches to dataset curation \cite{PODRYABINKIN2017171, zaverkin2024uncertainty,ANI_Al,kulichenko2023uncertainty}. Uncertainty quantification (UQ) of MLIAPs has therefore been the subject of extensive prior studies, both for classical IAPs \cite{kurniawan2022bayesian,hegde2022bayesian,hegde2024bayesian,longbottom2019uncertainty} and more recent MLIAPs \cite{Zhu2023,tan2023single,hu2022robust,busk2023graph,bartok2022improved,best2024uncertainty,zaverkin2024uncertainty}. 

%\tom{Tom to Danny: add a line on the dataset generation? how is the following for you?}
{The vast majority of existing UQ schemes (not just for MLIAPs) 
tacitly ignore uncertainties due to \textit{misspecification}, or model 
imperfection, i.e. the idea that no one choice of model parameters can exactly 
match all training data. Misspecification affects both finite capacity models and deep learning approaches with finite training resources
~\cite{lahlou2021deup,psaros2023uncertainty}.
It is however known that the loss (an upper bound to the true 
generalization error\cite{masegosa2020learning}) is only sensitive to epistemic 
(data-dependent) or aleatoric (intrinsic) uncertainties. Commonly used loss-based schemes can
therefore significantly underestimate parameter uncertainties and model errors.

We have recently introduced a misspecification-aware UQ framework to resolve 
this issue\cite{swinburne2025}, which is summarised below.}
%\tom{Tom to Danny: more detailed summary?}
{In this paper, we use this framework to demonstrate the quantification of misspecification uncertainties on MLIAP parameters and their propagation\cite{imbalzano2021uncertainty,maliyov2024impl_diff} to material properties of physical interest. We present extensive tests of property predictions against 
brute force DFT calculations on a diverse set of physical properties which were not explicitly included in the training data}.
{Our main result is that propagated misspecification uncertainties 
provide an efficient and robust means to assign informative error bars on 
simulated material properties. We show that in all of the diverse test cases 
considered, the misspecification prediction bounds contain the "true" answer, 
i.e. that calculated \textit{a posteriori} with the same DFT engine, and that the distribution of predictions offers a slightly conservative, but nonetheless generally quantitative representation of the actual errors.}

%\tom{Rework this I suppose}
This paper is organized as follows. After reviewing existing approaches for 
MLIAP UQ in \ref{sec:UQ}, section \ref{sec:methods} summarizes the UQ and ML 
methodologies leveraged in this work. Section \ref{sec:results} then presents 
an extensive characterization of the performance of our UQ approach on a broad 
set of tasks that relate to the prediction of perfect crystal and defect 
properties. Whilst most of the error propagation is achieved through resampling, i.e. 
repeating simulations with resampled parameters, we also test our recent implicit differentiation scheme \cite{maliyov2024impl_diff} in section 
\ref{sec:impl}. The accuracy of this method demonstrates how misspecification 
uncertainties in the interatomic potential can be efficiently propagated to 
simulation results of interest in a multiscale modeling workflow. In anticipation
of future work, we finally demonstrate how POPS can be used to predict and bound errors from recent foundational, or universal, message passing neural network 
MLIAPs \cite{batatia2023foundation,deng_2023_chgnet}. Perspectives for the method are then discussed in Sec.\ \ref{sec:conclusion}.

\subsection{Uncertainty Quantification for MLIAPs\label{sec:UQ}}
%Tom, can you review this section and add references as appropriate?
Approaches for UQ on MLIAP predictions have differed based on the model architecture employed and the goal of the UQ task. 
For active learning schemes the primary goal is uncertainty \textit{qualification}, i.e. a classification of whether an individual 
force or energy evaluation is trustworthy. If not, 
new \textit{ab initio} reference data is typically required to either 
refine or directly replace model predictions\cite{Zhenwei-2015,PODRYABINKIN2017171, zaverkin2024uncertainty,ANI_Al,kulichenko2023uncertainty}.

{MLIAPs that rely on Gaussian Processes Regression \cite{bartok2010gaussian,vandermause2020fly,bartok2022improved} posses an intrinsic uncertainty metric through the posterior variance, which is not typically a quantitative 
prediction but is ideal for use in active learning schemes. 
However, to provide robust uncertainty quantification 
on physical properties of interest, the MLIAP uncertainty
must be quantified and propagated to the results of any simulation.
Uncertainty propagation is challenging due to the strong correlations inherent
to most simulation data, whether e.g. a trajectory average or a formation 
energy. As a result, it is in general challenging to propagate the uncertainty 
from Gaussian Processes Regression; the most efficient approach is
typically estimating or sampling uncertainty on MLIAP parameters,
as then propagation reduces to repeating simulations with sampled 
parameters\cite{imbalzano2021uncertainty,lu2023uncertaintyestimates} or evaluating gradients of simulation results with respect to potential parameters\cite{maliyov2024impl_diff,blondel2022efficient}.}

{MLIAPs based on neural networks (NN) typically 
employ ensemble (query-by-committee) approaches \cite{seung1992query,artrith2012high,zaverkin2021exploration,ANI_Al,Zhu2023,schran2020committee,smith2018less,imbalzano2021uncertainty,kellner2024uncertainty,lakshminarayanan2017simple,lu2023uncertaintyestimates},
where multiple models are trained in different conditions (initial weights, hyperparameters, subsets of the training data). In many cases the ensemble only adjusts a subset of model weights for computational efficiency\cite{kellner2024uncertainty}. The result is an effective sample of plausible parameter values, which can be then propagated to properties either through direct resampling (rerunning simulations) or reweighting schemes\cite{imbalzano2021uncertainty}. UQ metrics then follow from the statistical properties of ensemble predictions.}

{A key strength of the ensemble approach is its simplicity, for only a mild increase in training cost (especially if only a subset of model weights are ensembled\cite{kellner2024uncertainty}), resulting in broad adoption 
in gauging uncertainty in neural network models, both in atomic simulation\cite{imbalzano2021uncertainty,kellner2024uncertainty,lu2023uncertaintyestimates} and more widely\cite{tripathy2018deep,lakshminarayanan2017simple}. However, ensemble approaches are a form of bagging predictor\cite{breiman1996bagging},
which are known to underestimate model errors\cite{kahle2022quality}
for learning problems with weak aleatoric noise, as is the case for IAP models\cite{lu2023uncertaintyestimates}. In practice, ensemble approaches typically require multimodal model loss functions 
to return appreciable ensemble variance, without any theoretical 
guarantees that ensemble methods produce robust or predictive 
bounds on errors. As a result, error predictions typically 
require some form of calibration\cite{imbalzano2021uncertainty} to be quantitative, as they are typically overconfident\cite{lu2023uncertaintyestimates}.
Within a framework of calibration-enabled error prediction, conformal UQ methods have also been applied to MLIAP errors\cite{best2024uncertainty}.}

% Do these advanced Bayesian NN approaches suffer from the same issue dealing with misspecification errors? Yes, as they remain based on loss minimization
For conventional IAPs and MLIAP that rely on linear ML architectures combined with strongly non-linear features, UQ approaches have traditionally relied on Bayesian regression \cite{bishop2003bayesian} to quantify parametric uncertainties \cite{frederiksen2004bayesian,hegde2022bayesian,kurniawan2022bayesian,hegde2024bayesian,williams2024active}, which can be extended to Bayesian NN \cite{goan2020bayesian}. In our recent work\cite{swinburne2025}, discussed in more detail below, we address a known shortcoming of all Bayesian regression which minimize some expected loss, irrespective of model architecture: 
the expected loss provably ignores uncertainty due to 
\textit{misspecification}, or imperfection, 
where no one choice of model parameters can 
perfectly predict training data. The vast majority of 
regression schemes target the loss and thus significantly underestimate parameter uncertainties (i.e. model errors) in the large-data, low-noise limit of interest for MLIAP fitting\cite{masegosa2020learning,bayarri2007framework,sargsyan2019embedded}. In this limit, misspecification errors dominate, leading to bias and underestimation of uncertainties\cite{swinburne2025,masegosa2020learning} if misspecification is ignored. 

Whilst a small number of misspecification-aware Bayesian regression methods exist\cite{sargsyan2019embedded,masegosa2020learning,kato2022view,morningstar2022pacm,Kleijn-Vaart-2006,Kleijn-Vaart-2012}, they are only 
numerically stable in the regime of appreciable aleatoric uncertainty, whilst 
MLIAP models are fit to near-deterministic electronic structure calculations 
with vanishing aleatoric error\cite{swinburne2025}. Our recent 
scheme\cite{swinburne2025} is thus uniquely able to estimate misspecification uncertainties for MLIAP fitting. 

\section{\label{sec:methods}Methods}
\subsection{Misspecification-aware Bayesian regression for MLIAP fitting}
In the following, we demonstrate the effectiveness a recently-introduced misspecification-aware UQ method to describe the uncertainties inherent to MLIAPs. To summarize the above, this method specifically targets the aforementioned regime where:
\begin{enumerate}[nolistsep]
    \item The reference data (here DFT energies and forces) is near-deterministic, i.e., it exhibits vanishing aleatoric errors
    \item The ML model is misspecified, i.e., no single choice of the free parameters can reproduce all reference data exactly
    \item The model is underparameterized, i.e., the amount of training data significantly exceeds the number of trainable parameters
\end{enumerate}

In the context of MLIAPs, condition 1 reflects the near-deterministic nature of well-converged quantum calculations, where repeated calculations with the same inputs result in the same output. While some MLIAP formalism provide completeness guarantees in some limit, practical accuracy/computational cost tradeoffs \cite{xie2023ultra} commonly results in the use of misspecified models where conditions 2 and 3 are met. In this regime, uncertainties on the predictions derived from the MLIAPs do not stem from the intrinsically noisy nature of the data or from insufficient amount of training data, but are dominated by the misspecified nature of the ML model.

In the following, we will demonstrate that this approach provides i) reliable estimates of point-wise energy and force errors, ii) reliable bounds on maximal errors, and iii) reliable errors estimates on a large number of non-point-wise complex properties (e.g., formation energies, energy barriers, etc.),
which enables a thorough characterization of the uncertainties obtained by MLIAPs at a very affordable computational cost. This enables a systematic approach to the evaluation of the predictability of the simulation results that goes beyond what would be possible using point-wise average metrics alone.

\subsection{\label{sec:cube}POPS-hypercube \textit{ansatz} for linear models}
For completeness, this section summarizes the key details of our scheme to quantify misspecification uncertainties. 
We refer the reader to Ref. \citep{swinburne2025} for a detailed 
presentation. An open source implementation, following the Scikit-learn \texttt{linear\_model} API\cite{sklearn}, is available on GitHub at \url{https://github.com/tomswinburne/POPSRegression.git}

Our goal is to determine a posterior distribution $\pi({\bf\Theta})$ of parameters for some MLIAP $\mathcal{M}({\bf X};{\bf\Theta})$, which aims
to approximate some DFT ground truth $\mathcal{E}({\bf X})$. In the following derivation (but not in the numerical experiments that follow)
we only consider energies for brevity, with the extension to forces trivial. 
From a Bayesian perspective, the near-deterministic nature of $\mathcal{E}$ 
is manifest in the sharp conditional distribution of output ${\rm Y}$ (here a scalar energy) given an input ${\bf X}$, reading
\begin{equation}
    \rho_\mathcal{E}({\rm Y}|{\bf X})=
    \exp(\|\mathcal{E}({\bf X})-{\rm Y})\|^2/\epsilon^2)/\sqrt{\pi\epsilon^2}
\end{equation}
which limits to a delta function as $\epsilon\to0$. 
Bayesian regression aims to find the distribution of model parameters 
$\pi({\bf\Theta})$ to minimize the cross entropy between $\rho_\mathcal{E}({\rm Y}|{\bf X})$ and the model distribution, which writes 
\begin{eqnarray}
    \rho_\mathcal{M}({\rm Y}|{\bf X},\pi)
    =
    \int
    \exp(-\|\mathcal{M}({\bf X};{\bf\Theta})-{\rm Y}\|^2/\epsilon^2)
    \frac{{\rm d}\pi({\bf\Theta})}{\sqrt{\pi\epsilon^2}},
\end{eqnarray}
where ${\rm d}\pi({\bf\Theta})=\pi({\bf\Theta}){\rm d}{\bf\Theta}$.
The cross entropy between $\rho_\mathcal{M}({\rm Y}|{\bf X},\pi)$ 
and $ \rho_\mathcal{E}({\rm Y}|{\bf X})$ 
is known as the \textit{generalization error}, here 
$\mathcal{G}[\pi]$, reading (see \cite{swinburne2025} 
for a full derivation)
\begin{equation}
    \mathcal{G}[\pi]
    =
    %\mathcal{G}_0
    -\left\langle 
        \ln
        \left|
        \int
        \exp(-\|\mathcal{M}({\bf X};{\bf\Theta})-\mathcal{E}({\bf X})\|^2/\epsilon^2)
        {\rm d}\pi({\bf\Theta})\right|
    \right\rangle,
    \nonumber
\end{equation}
where $\langle\dots\rangle$ denotes an average over a formally infinite quantity of training data, potentially with a normalized positive
weighting $w({\bf X})$. 
Minimization of $\mathcal{G}[\pi]$ is extremely challenging due to the poor 
conditioning of the logarithmic term, and also does not have any means to 
incorporate epistemic (finite data) uncertainties. However, it is clear that 
unless a single value of $\bf\Theta$ can produce perfect predictions, 
$\pi({\bf\Theta})$ is required to have finite width, which is precisely the misspecification uncertainty we wish to estimate. 
As $\mathcal{G}[\pi]$ is numerically intractable, the vast 
majority of regression techniques employ the Jensen inequality $-\langle\ln x\rangle\leq-\ln\langle x\rangle$ for convex functions to define $\mathcal{L}[\pi]$, the \textit{expected loss} or log likelihood through 
\begin{equation}
    \mathcal{G}[\pi]
\leq\mathcal{L}[\pi]
    =
    \frac{1}
    {\epsilon^2}
    \int
    \left\langle
        \|\mathcal{M}({\bf X};{\bf\Theta})-\mathcal{E}({\bf X})\|^2
    \right\rangle
    {\rm d}\pi({\bf\Theta}).
    \nonumber
\end{equation}
It is clear that $\mathcal{L}[\pi]$ is minimized by a sharp distribution around the global loss minimizer 
\begin{equation}
    {\bf\Theta}^*\in\arg\min\left\langle\|\mathcal{M}({\bf X};{\bf\Theta})-\mathcal{E}({\bf X})\|^2\right\rangle,    
\end{equation}
such that $\pi^*_\mathcal{L}({\bf\Theta})=\delta({\bf\Theta}-{\bf\Theta}^*)$. 
This important result shows that loss minimization ignores misspecification uncertainties, which as discussed above are dominant for MLIAPs.
The connection to Bayesian inference at finite data (i.e. with epistemic uncertainties) was made in \cite{germain2016pac}, using PAC-Bayes analysis\cite{hoeffding1994probability,alquier2021user} to show that
\begin{equation}
\mathcal{L}[\pi]
    \leq
    C+
    \int
    \left[
    \frac{\sigma^2_N({\bf\Theta})}{\epsilon^2}
    +
    \frac{1}{N}\ln\frac{\pi({\bf\Theta})}{\pi_0({\bf\Theta})}
    \right]
    \pi({\bf\Theta})
    {\rm d}{\bf\Theta},
\end{equation}
where $\sigma^2_N({\bf\Theta})=\sum_i w_i\|\mathcal{M}({\bf X}_i;{\bf\Theta})-\mathcal{E}({\bf X}_i)\|^2/N$ is the average squared error over the $N$ training points, $C$ is a constant\cite{swinburne2025,germain2016pac} and $\pi_0({\bf\Theta})$ is some \textit{prior} distribution. 
It is simple to show this upper bound is minimized by the well-known 
posterior from Bayesian inference
\begin{equation}
    \pi^*_N({\bf\Theta})=\pi_0({\bf\Theta})\exp[-N{\sigma^2_N({\bf\Theta})}/{\epsilon^2}].
\end{equation}
In the large data limit $N\to\infty$, application of steepest decents recovers the sharp distribution $\pi^*_\mathcal{L}({\bf\Theta})$, again showing the inability to capture misspecification uncertainty. 

To find an approximate minimizer of $\mathcal{G}$, our approach\cite{swinburne2025} defines \textit{pointwise optimal parameter sets} (POPS) for each training point $\bf X$, being the set of all model parameters where that particular training point is exactly matched, i.e. all ${\bf\Theta}$ such that $\mathcal{M}({\bf X};{\bf\Theta})=\mathcal{E}({\bf X})$ at $\bf X$. 
In Ref.\ \cite{swinburne2025} we show that any posterior distribution 
$\pi$ which minimizes the generalization error must have mass in every POPS in 
the training set. For misspecified models, the mutual intersection of all POPS 
is empty, enforcing a finite parameter uncertainty. Our POPS regression algorithm first finds the parameter $\boldsymbol{\Theta}^*_{\bf X}$ that 
minimizes the global loss conditional on lying in the POPS of ${\bf X}$.
This produces an ensemble of $N$ parameter values clustered around the global loss minimizer ${\bf\Theta}^*_\mathcal{L}$. The final parameter posterior distribution $\pi^*_\mathcal{H}$ is then taken a uniform distribution
over the minimal hypercube $\mathcal{H}$ in parameter space that encompasses all of the $N$ POPS-constrained loss minimizers. 
For a model of $P$ parameters, the POPS-hypercube posterior 
can then be resampled for only $\mathcal{O}(P)$ computational effort and is thus a highly efficient manner to capture the dominant uncertainty in interatomic potentials trained on large datasets. Our open source implementation incurs a minimial overhead of around $\times2$ over 
Bayesian ridge regression as implemented in Sci-Kit learn's \texttt{linear\_model.BayesianRidge}.

\subsection{\label{sec:training} Interatomic potential training }
We consider Machine Learned interatomic potentials in the familly of the Spectral Neighbor Analysis Potential (SNAP) \cite{thompson2015spectral}, more specifically of the quadratic SNAP (qSNAP) type \cite{wood2018extending}. SNAP potentials describe the local environment around an atom $i$ in terms of invariants of a spherical harmonics expansion of the local atomic density,  the so-called bispectrum components denoted $\{B^i_k\}$. Under qSNAP, the corresponding atomic energy is expanded to second order in bispectrum components, i.e.,
\begin{equation}
    E^i_{\rm SNAP}=\boldsymbol{\beta \cdot B^i + \frac{1}{2} (B^i)^T \cdot \alpha \cdot B^i},
\end{equation}
where $\boldsymbol{\beta}$ and $\boldsymbol{\alpha}$ are vectors and matrices of adjustable coefficients, respectively.  
For simplicity we collate the linear and quadratic terms into a single parameter vector 
$\boldsymbol{\Theta}$ and descriptor vector ${\bf D}^i$,
giving the atomic energy as 
\begin{equation}
E^i_{\rm SNAP}
    =
    \boldsymbol{\Theta}\cdot{\bf D}^i,
    \label{eq:SNAP}
\end{equation}
The total energy of a configuration of atoms is then defined as the sum of the per atom energies,
\begin{equation}
    \mathcal{M}({\bf X};{\bf\Theta})
    =\sum^N_{i=1} E^i_{\rm SNAP} =\sum^N_{i=1} \boldsymbol{\Theta}\cdot{\bf D}^i
    \label{eq:SNAP_energy}
\end{equation}
and the atomic forces as the gradient of Eq.\ \ref{eq:SNAP_energy} with respect to atomic coordinates.

Training a qSNAP model therefore corresponds to solving a (potentially weighted) linear least-square problem with unknowns $\boldsymbol{\beta}$ and $\boldsymbol{\alpha}$ so as to minimize squared (total) energy and (atomic) force residuals against reference quantum calculations.  

The reference dataset was here obtained using a diverse-by-construction generation technique introduced in Refs.\ \cite{karabin2020entropy,montes2022training} and generalized in Ref.\ \cite{subramanyam2024information}. This method creates atomic configurations so as to specifically maximize the information entropy of the bispectrum component distribution, resulting in very broad coverage of feature space. The dataset considered here was introduced in Ref.\ \cite{subramanyam2024information}, and was rescaled to the interatomic spacing of tungsten. The data was partitioned into a training set containing 7000 energies and 122,853 force components and a testing set containing 3000 energies and 53,493 force components.

Since the properties of lower energy structures are often the target of practical investigations, individual energies and forces were weighted to give more importance to near-equilibrium configurations, following:
\begin{equation}
\begin{aligned}
    w_\mathrm{energy} & \propto & \exp(- E_\mathrm{ref}/a) \\
    w_\mathrm{force} & \propto & \exp(- |F_\mathrm{ref}|/b)
\end{aligned}
\end{equation}
with $a=2$ eV and $b=50$ eV/\AA. The energy and force weights are normalized so that their respective sums over the training set are equal.

The potentials considered here were not fine-tuned nor the hyper-parameters (like $a$ and $b$, the cutoff radius, etc.) optimized, as this potential was designed to serve as an assessment of the performance of the UQ procedure, not to generate a production-optimal model.

\subsection{\label{sec:ensemble} UQ ensemble}

\begin{figure}[!htb]
\includegraphics[width=0.49\columnwidth]{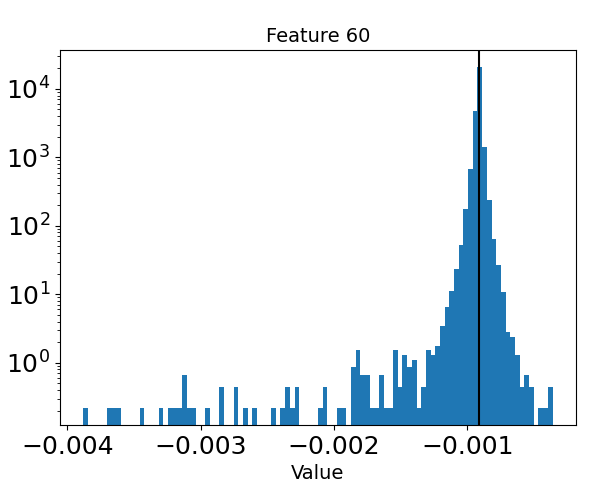}
\includegraphics[width=0.49\columnwidth]{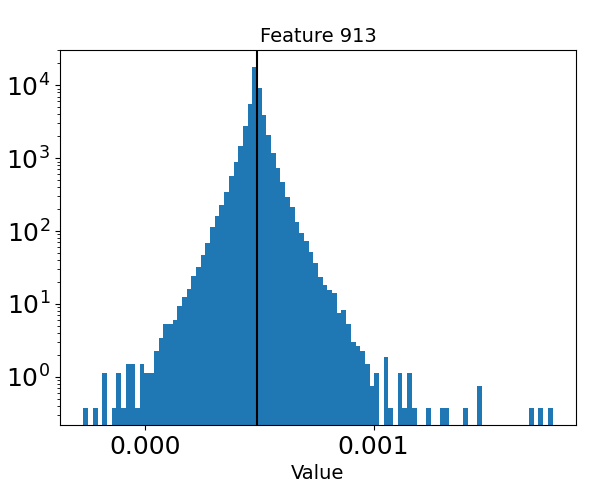}
\caption{\label{fig:histo} Distribution of regression coefficients over the loss-minimizing POPS ensemble $\pi^*_E$ . The corresponding MLE coefficients are shown by the black vertical line.
}
\end{figure}

\begin{figure}[!htb]
\includegraphics[width=0.98\columnwidth]{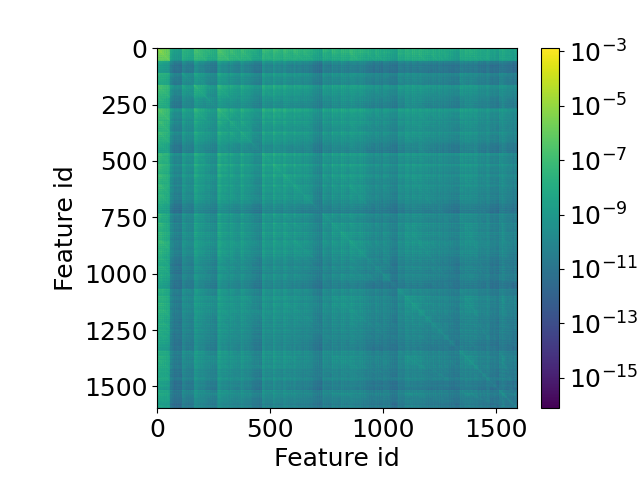}
\caption{\label{fig:covariance} Absolute value of the covariance of the regression coefficients over the loss-minimizing POPS ensemble $\pi^*_E$.
}
\end{figure}

The weighted least squares solution will be referred to as the MLE solution. 
In a first stage, a loss-minimizing POPS ensemble $\pi^*_E$ containing 129,853 models was  generated according to the procedure described in Sec.\ \ref{sec:cube}. 
The distribution of selected regression coefficients in $\pi^*_E$ reported in Fig.\ \ref{fig:histo} shows a strongly non-gaussian behavior and the occasional presence of very fat and asymmetric tails (e.g., for Feature 60). Furthermore, as shown in Fig.\ \ref{fig:covariance}, the coefficients over the ensemble are correlated with each other following a complex pattern that reflects the physical definition of the features, the product structure of Eq.\ \ref{eq:SNAP} (which can be expected to introduce correlations between regression coefficients), and their relative importance in the regression task. 

An ensemble $\pi^*_\mathcal{H}$ of 500 models was then uniformly resampled from the  hyper-cube bounding $\pi^*_E$, a procedure which was previously shown to provide very good statistical error estimates at a small computational cost. Unless otherwise noted, the UQ ensemble results reported below are generated from $\pi^*_\mathcal{H}$.

\subsection{\label{sec:validation_suite} Validation suite }
The ability of the ensemble $\pi^*_\mathcal{H}$ to characterize the uncertainty on the predictions of the MLE model is assessed using a comprehensive validation suite of properties that are often of interest in practical applications of MLIAPs, including perfect crystal properties, defects, and energy barriers. 
Note that none of the validation properties were explicitly included in the training data, which was generated without any input from domain experts according to the procedure described in Ref. \cite{subramanyam2024information}, and therefore can be considered as an assessment of the UQ procedure on genuinely unseen test data.

\section{\label{sec:results}Results}
%{\bf Aparna, please describe the validation suite corresponding to the figures below. Just a few sentences per property.}

\subsection{\label{sec:pointwise} Pointwise properties}
\begin{figure}[!htb]
\includegraphics[width=0.98\columnwidth]{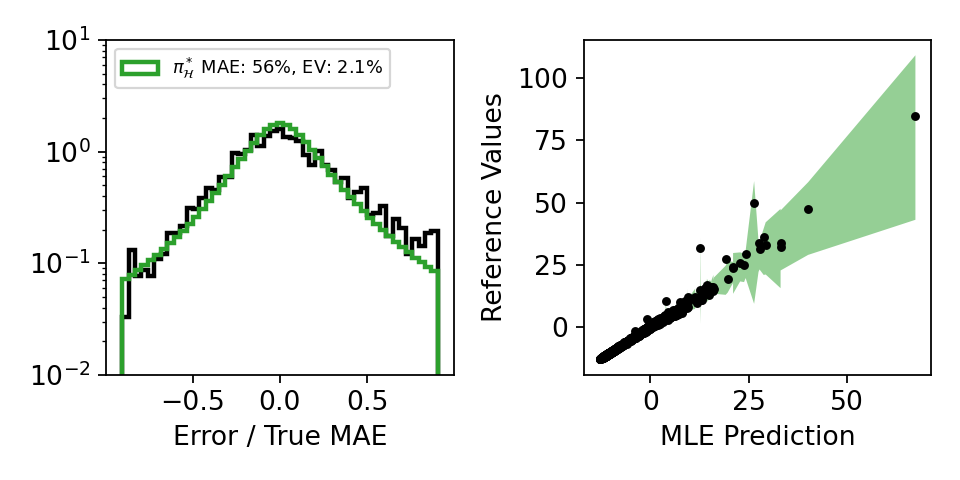}
\includegraphics[width=0.98\columnwidth]{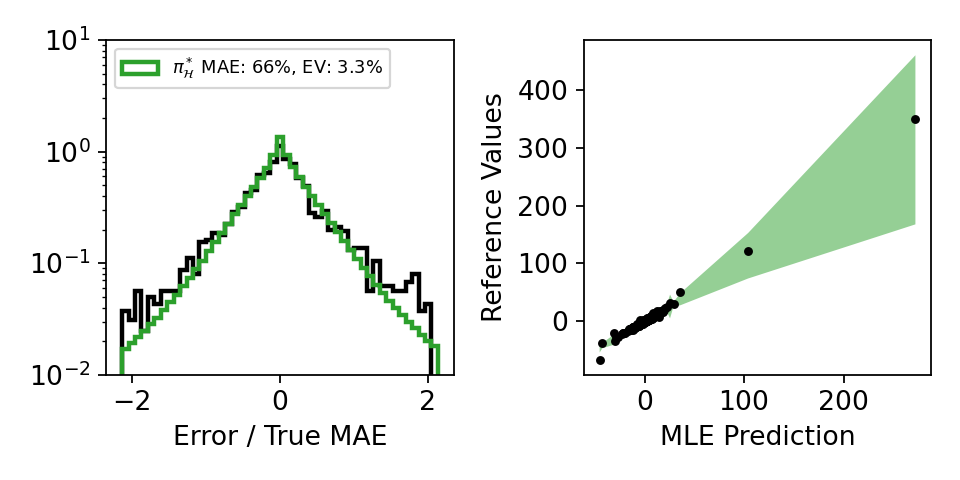}
\caption{\label{fig:pointwise} 
Characterization of the statistics of pointwise errors obtained from $\pi^*_\mathcal{H}$.
Top: UQ on energies; Bottom: UQ on forces.
Left: distribution of test errors for the MLE against the reference data (black) and 
from $\pi^*_\mathcal{H}$ \textit{ansatz} against the MLE (green). MAE: mean absolute error relative to the minimum loss solution. EV: envelope violation, fraction of points lying outside of the max/min bound. Right: parity plot of actual vs MLE predicted energies. Shaded areas show the min/max range of predictions over all members of $\pi^*_\mathcal{H}$.
}
\end{figure}

In keeping with the traditional ML literature, the most common approach to characterizing the performance of MLIAPs is through point-wise error metrics measured on a testing set that is nominally independent of the training set. Predicting the distribution of errors incurred by the MLIAP is therefore a natural objective. Fig.\ \ref{fig:pointwise} a) reports the distribution of the ratio of actual point-wise test errors to the difference between the MLE and UQ ensemble models. The results 
demonstrates that the overall error distributions from the MLE is extremely well captured by the resampled POPS ensemble $\pi^*_\mathcal{H}$. This shows that the deviation between the predictions of the MLE and that of individual samples from the model ensemble provide a representative statistical estimate of the actual difference between the MLE and (often unknown) exact reference value. The ensemble also provides excellent bounds on predictions: maximal and minimal predictions of an ensemble of 500 models sampled from $\pi^*_\mathcal{H}$ fails to bound the actual reference energies and forces in only 2.1\% and 3.3\% of the case, respectively.
Furthermore, the bounds provided by the model ensemble capture very specific features of individual predictions. E.g., in addition to capturing the generic increase in error with increasing energy or forces that results from the reweighting scheme used to train the MLIAP, "outlier" points with unusually large errors compared to their neighboring peers are very accurately captured (c.f., the outlier points in Fig.\ \ref{fig:pointwise}). 
These results clearly shows that the UQ ensemble does not only capture average error behavior, but closely resolves high uncertainty regions that result from particularly detrimental combinations of test point and intrinsic model limitations. The ability to confidently bound predictions is also a powerful feature that can be used to easily propagate worst-case scenarios to more complex quantities, as will be shown in the following.

%\begin{figure}[!htb]
%\includegraphics[width=0.49\columnwidth]{figures/energy-bounds.png}
%\includegraphics[width=0.49\columnwidth]{figures/energy-bounds-zoom.png}
%\includegraphics[width=0.49\columnwidth]{figures/force-bounds.png}
%\includegraphics[width=0.49\columnwidth]{figures/force-bounds-zoom.png}
%%\caption{\label{fig:pointwise-bounds} Top row: parity plots for energy predictions. 
%Bottom row: parity plots for force predictions.
%Blue dots: MLE predictions; vertical bars: max/min range from the ensemble.}
%\end{figure}

\subsection{\label{sec:perfect} Perfect crystal properties}

\begin{figure}[!htb]
\includegraphics[width=0.49\columnwidth]{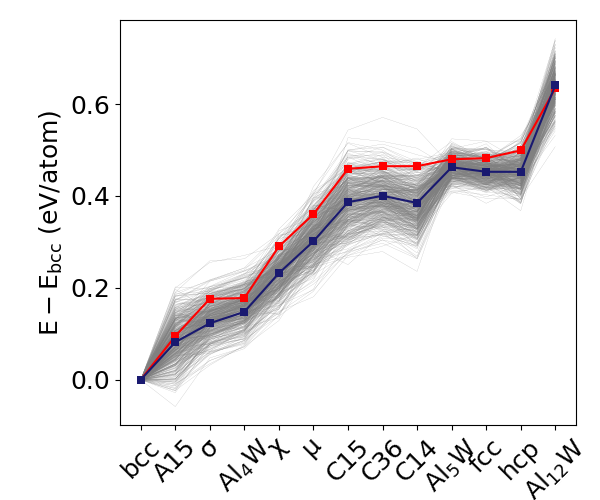}
\includegraphics[width=0.49\columnwidth]{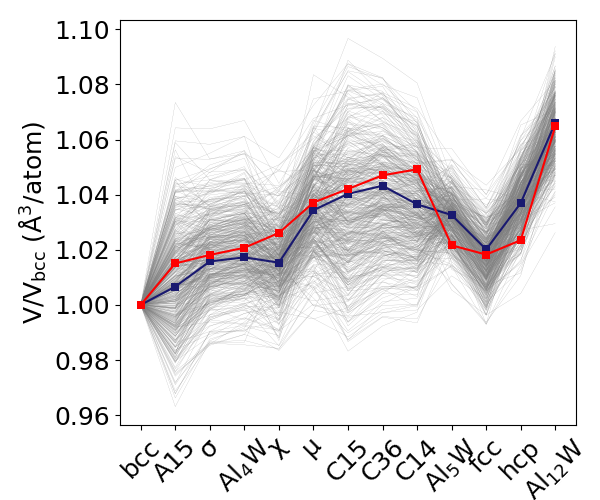}
\includegraphics[width=0.49\columnwidth]{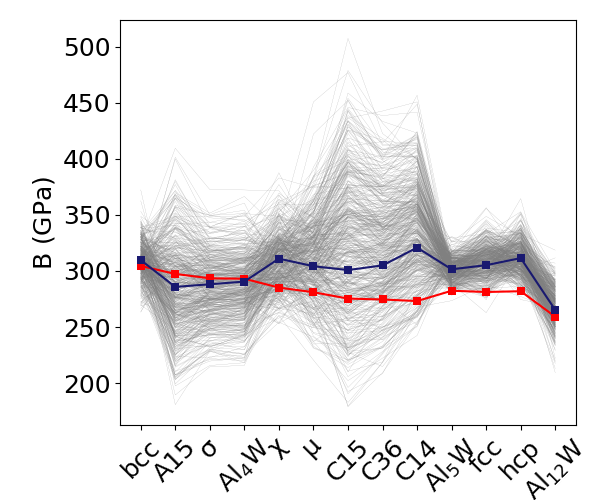}

\caption{\label{fig:phases} Left: Equilibrium formation energy for different crystal structures relative to the BCC phase. Right: ratio of equilibrium volumes to the BCC phase; Bottom: bulk moduli. MLE predictions are shown in blue, DFT reference values in red, and ensemble predictions in grey. }
\end{figure}

\begin{table}
    \centering
    \caption{UQ statistics for the crystal formation energy of different crystal phases relative to the BCC phase (c.f., Fig.\ \ref{fig:phases}). MLE error corresponds to the errors of the MLE relative to the DFT reference, Std is the standard deviation of the predictions of $\pi^*_\mathcal{H}$, Low Error is the difference between the smallest prediction in $\pi^*_\mathcal{H}$ and the DFT reference, and High Error is the difference between the largest prediction in $\pi^*_\mathcal{H}$ and the DFT reference. A negative Low Error and a positive High Error indicate that the predictions from the UQ ensemble bracket the reference value. }
    \label{tab:energies}
    \begin{tabular}{ccccc}
        \toprule
        Structure & MLE Error & Std & Low Error & High Error \\
        \midrule
        bcc & $0.00\times 10^{ 0 }$ & $0.00\times 10^{ 0 }$ & $0.00\times 10^{ 0 }$ & $0.00\times 10^{ 0 }$ \\
        A15 & $-1.30\times 10^{ -2 }$ & $4.60\times 10^{ -2 }$ & $-1.53\times 10^{ -1 }$ & $1.07\times 10^{ -1 }$ \\
        $\mathrm{\sigma}$ & $-5.33\times 10^{ -2 }$ & $3.49\times 10^{ -2 }$ & $-1.45\times 10^{ -1 }$ & $8.25\times 10^{ -2 }$ \\
        Al$\mathrm{_4}$W & $-3.00\times 10^{ -2 }$ & $3.42\times 10^{ -2 }$ & $-1.10\times 10^{ -1 }$ & $9.30\times 10^{ -2 }$ \\
        $\mathrm{\chi}$ & $-5.81\times 10^{ -2 }$ & $3.58\times 10^{ -2 }$ & $-1.61\times 10^{ -1 }$ & $3.76\times 10^{ -2 }$ \\
        $\mathrm{\mu}$ & $-5.92\times 10^{ -2 }$ & $4.07\times 10^{ -2 }$ & $-1.80\times 10^{ -1 }$ & $6.03\times 10^{ -2 }$ \\
        C15 & $-7.25\times 10^{ -2 }$ & $5.03\times 10^{ -2 }$ & $-2.09\times 10^{ -1 }$ & $8.43\times 10^{ -2 }$ \\
        C36 & $-6.43\times 10^{ -2 }$ & $4.25\times 10^{ -2 }$ & $-1.86\times 10^{ -1 }$ & $1.06\times 10^{ -1 }$ \\
        C14 & $-8.01\times 10^{ -2 }$ & $4.46\times 10^{ -2 }$ & $-2.29\times 10^{ -1 }$ & $8.08\times 10^{ -2 }$ \\
        Al$\mathrm{_5}$W & $-1.74\times 10^{ -2 }$ & $2.08\times 10^{ -2 }$ & $-7.41\times 10^{ -2 }$ & $4.44\times 10^{ -2 }$ \\
        fcc & $-2.98\times 10^{ -2 }$ & $2.15\times 10^{ -2 }$ & $-9.83\times 10^{ -2 }$ & $3.71\times 10^{ -2 }$ \\
        hcp & $-4.67\times 10^{ -2 }$ & $2.84\times 10^{ -2 }$ & $-1.31\times 10^{ -1 }$ & $3.24\times 10^{ -2 }$ \\
        Al$\mathrm{_{12}}$W & $6.69\times 10^{ -3 }$ & $3.87\times 10^{ -2 }$ & $-1.27\times 10^{ -1 }$ & $1.09\times 10^{ -1 }$ \\
        \midrule
        Mean $\frac{\mathrm{Error}}{\mathrm{Std}}$ & 1.21 & \multicolumn{2}{c}{Mean $\frac{\mathrm{Error}}{\mathrm{High-Low}}$} & 0.19 \\
        \bottomrule
    \end{tabular}
\end{table}

\begin{table}
    \centering
    \caption{UQ statistics for the crystal formation volume of different crystal phases relative to the volume of the BCC phase (c.f., Fig.\ \ref{fig:phases}). MLE error corresponds to the errors of the MLE relative to the DFT reference, Std is the standard deviation of the predictions of $\pi^*_\mathcal{H}$, Low Error is the difference between the smallest prediction in $\pi^*_\mathcal{H}$ and the DFT reference, and High Error is the difference between the largest prediction in $\pi^*_\mathcal{H}$ and the DFT reference. A negative Low Error and a positive High Error indicate that the predictions from the UQ ensemble bracket the reference value. }
    \label{tab:volume}
    \begin{tabular}{ccccc}
        \toprule
        Structure & MLE Error & Std & Low Error & High Error \\
        \midrule
        bcc & $0.00\times 10^{ 0 }$ & $0.00\times 10^{ 0 }$ & $0.00\times 10^{ 0 }$ & $0.00\times 10^{ 0 }$ \\
        A15 & $-8.41\times 10^{ -3 }$ & $1.77\times 10^{ -2 }$ & $-5.19\times 10^{ -2 }$ & $5.84\times 10^{ -2 }$ \\
        $\mathrm{\sigma}$ & $-2.36\times 10^{ -3 }$ & $1.35\times 10^{ -2 }$ & $-3.28\times 10^{ -2 }$ & $4.57\times 10^{ -2 }$ \\
        Al$\mathrm{_4}$W & $-3.48\times 10^{ -3 }$ & $1.35\times 10^{ -2 }$ & $-3.52\times 10^{ -2 }$ & $4.62\times 10^{ -2 }$ \\
        $\mathrm{\chi}$ & $-1.08\times 10^{ -2 }$ & $1.19\times 10^{ -2 }$ & $-4.26\times 10^{ -2 }$ & $2.72\times 10^{ -2 }$ \\
        $\mathrm{\mu}$ & $-2.78\times 10^{ -3 }$ & $1.41\times 10^{ -2 }$ & $-4.23\times 10^{ -2 }$ & $4.64\times 10^{ -2 }$ \\
        C15 & $-1.78\times 10^{ -3 }$ & $2.08\times 10^{ -2 }$ & $-5.88\times 10^{ -2 }$ & $5.46\times 10^{ -2 }$ \\
        C36 & $-3.78\times 10^{ -3 }$ & $1.78\times 10^{ -2 }$ & $-5.46\times 10^{ -2 }$ & $4.24\times 10^{ -2 }$ \\
        C14 & $-1.27\times 10^{ -2 }$ & $1.53\times 10^{ -2 }$ & $-5.57\times 10^{ -2 }$ & $3.13\times 10^{ -2 }$ \\
        Al$\mathrm{_5}$W & $1.10\times 10^{ -2 }$ & $7.93\times 10^{ -3 }$ & $-1.62\times 10^{ -2 }$ & $3.52\times 10^{ -2 }$ \\
        fcc & $1.86\times 10^{ -3 }$ & $9.07\times 10^{ -3 }$ & $-2.54\times 10^{ -2 }$ & $2.52\times 10^{ -2 }$ \\
        hcp & $1.35\times 10^{ -2 }$ & $9.44\times 10^{ -3 }$ & $-1.93\times 10^{ -2 }$ & $4.33\times 10^{ -2 }$ \\
        Al$\mathrm{_{12}}$W & $1.34\times 10^{ -3 }$ & $1.05\times 10^{ -2 }$ & $-3.85\times 10^{ -2 }$ & $2.90\times 10^{ -2 }$ \\
        \midrule
        Mean $\frac{\mathrm{Error}}{\mathrm{Std}}$ & 0.52 & \multicolumn{2}{c}{Mean $\frac{\mathrm{Error}}{\mathrm{High-Low}}$} & 0.085 \\
        \bottomrule
    \end{tabular}
\end{table}

\begin{table}
    \centering
\caption{UQ statistics for the bulk moduli of different crystal phases (c.f., Fig.\ \ref{fig:phases}). MLE error corresponds to the errors of the MLE relative to the DFT reference, Std is the standard deviation of the predictions of $\pi^*_\mathcal{H}$, Low Error is the difference between the smallest prediction in $\pi^*_\mathcal{H}$ and the DFT reference, and High Error is the difference between the largest prediction in $\pi^*_\mathcal{H}$ and the DFT reference. A negative Low Error and a positive High Error indicate that the predictions from the UQ ensemble bracket the reference value. }    \label{tab:moduli}
    \begin{tabular}{ccccc}
        \toprule
        Structure & MLE Error & Std & Low Error & High Error \\
        \midrule
        bcc & $5.24\times 10^{ 0 }$ & $1.65\times 10^{ 1 }$ & $-4.12\times 10^{ 1 }$ & $6.79\times 10^{ 1 }$ \\
        A15 & $-1.16\times 10^{ 1 }$ & $3.95\times 10^{ 1 }$ & $-1.17\times 10^{ 2 }$ & $1.12\times 10^{ 2 }$ \\
        $\mathrm{\sigma}$ & $-5.30\times 10^{ 0 }$ & $2.74\times 10^{ 1 }$ & $-7.89\times 10^{ 1 }$ & $7.91\times 10^{ 1 }$ \\
        Al$\mathrm{_4}$W & $-2.55\times 10^{ 0 }$ & $2.75\times 10^{ 1 }$ & $-7.73\times 10^{ 1 }$ & $7.91\times 10^{ 1 }$ \\
        $\mathrm{\chi}$ & $2.58\times 10^{ 1 }$ & $2.32\times 10^{ 1 }$ & $-3.22\times 10^{ 1 }$ & $1.02\times 10^{ 2 }$ \\
        $\mathrm{\mu}$ & $2.32\times 10^{ 1 }$ & $3.25\times 10^{ 1 }$ & $-6.05\times 10^{ 1 }$ & $1.70\times 10^{ 2 }$ \\
        C15 & $2.54\times 10^{ 1 }$ & $5.85\times 10^{ 1 }$ & $-9.63\times 10^{ 1 }$ & $2.32\times 10^{ 2 }$ \\
        C36 & $3.04\times 10^{ 1 }$ & $4.59\times 10^{ 1 }$ & $-6.61\times 10^{ 1 }$ & $1.68\times 10^{ 2 }$ \\
        C14 & $4.76\times 10^{ 1 }$ & $4.09\times 10^{ 1 }$ & $-3.07\times 10^{ 1 }$ & $1.83\times 10^{ 2 }$ \\
        Al$\mathrm{_5}$W & $1.91\times 10^{ 1 }$ & $9.87\times 10^{ 0 }$ & $-8.85\times 10^{ 0 }$ & $4.85\times 10^{ 1 }$ \\
        fcc & $2.39\times 10^{ 1 }$ & $1.42\times 10^{ 1 }$ & $-1.84\times 10^{ 1 }$ & $7.51\times 10^{ 1 }$ \\
        hcp & $2.95\times 10^{ 1 }$ & $1.42\times 10^{ 1 }$ & $-7.47\times 10^{ 0 }$ & $8.26\times 10^{ 1 }$ \\
        Al$\mathrm{_{12}}$W & $5.97\times 10^{ 0 }$ & $1.66\times 10^{ 1 }$ & $-5.00\times 10^{ 1 }$ & $5.93\times 10^{ 1 }$ \\
        \midrule
        Mean $\frac{\mathrm{Error}}{\mathrm{Std}}$ & 0.89 & \multicolumn{2}{c}{Mean $\frac{\mathrm{Error}}{\mathrm{High-Low}}$} & 0.15 \\
        \bottomrule
    \end{tabular}
\end{table}

A second key class of properties of direct interest to applications is the quantification of the stability of different crystal structures. Fig.\ \ref{fig:phases} and Tables\ \ref{tab:energies} to \ref{tab:moduli} demonstrate that the UQ ensemble accurately captures the actual errors in formation energy, equilibrium volume, and bulk modulus over 13 different crystal structures that vary broadly in topology and unit cell sizes. These results are obtained by using atomistic configuration and simulation cells that were individually optimized under corresponding MLIAPs, in contrast to evaluating point-wise energies on the reference structures relaxed with DFT.

The results clearly show that the UQ ensemble accurately captures the uncertainty inherent to different phases, providing tightly distributed predictions where the actual errors are low and more diverse predictions where the actual errors are large, in addition to accurately bounding the actual predictions in all cases. 

%The analysis correctly captures the fact that formation energies are comparatively better captured than formation volumes and bulk moduli, perhaps reflecting the fact that the MLIAP was not explicitly trained to stresses. 
Tables\ \ref{tab:energies} to \ref{tab:moduli}  also show that the standard deviation of the UQ ensemble predictions provide a statistically representative indication of the magnitude of the actual errors, as the mean ratio of the MLE error to the standard deviation of the ensemble is close to 1, except for the formation volume where the ensemble overestimates the errors by about a factor of 2 on average. 
% Can we think of an explanation for this?

In all cases, the extreme values predicted by the ensemble bound the actual reference result, providing strong guarantees. 

\begin{figure}[!htb]
\includegraphics[width=0.49\columnwidth]{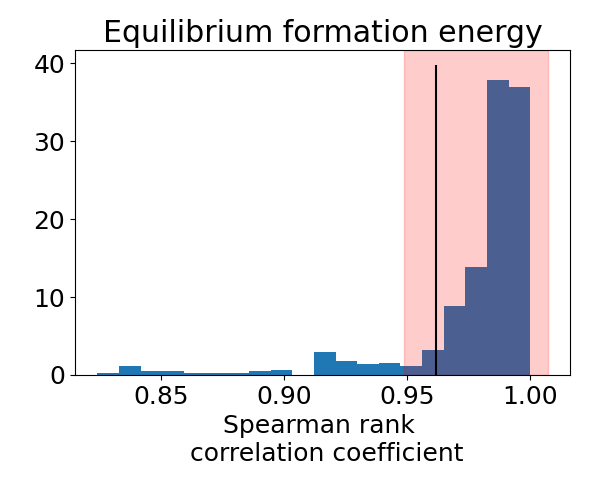}
\includegraphics[width=0.49\columnwidth]{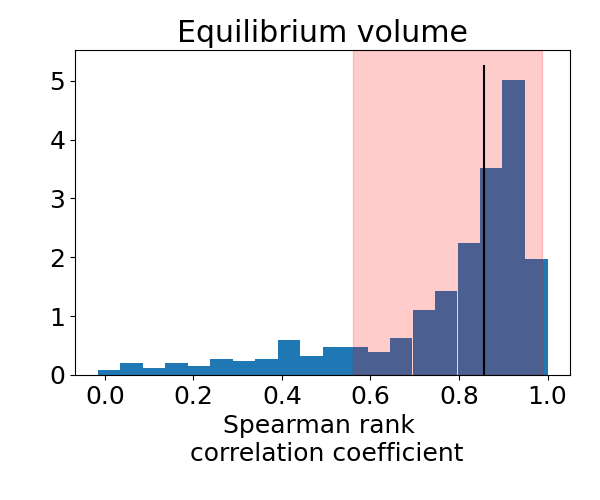}
\includegraphics[width=0.49\columnwidth]{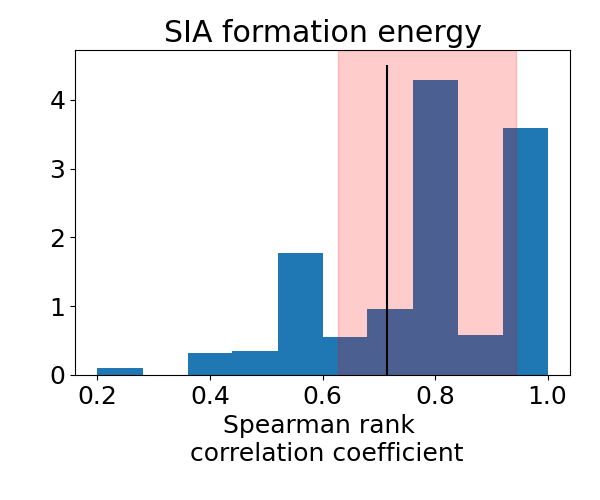}
\includegraphics[width=0.49\columnwidth]{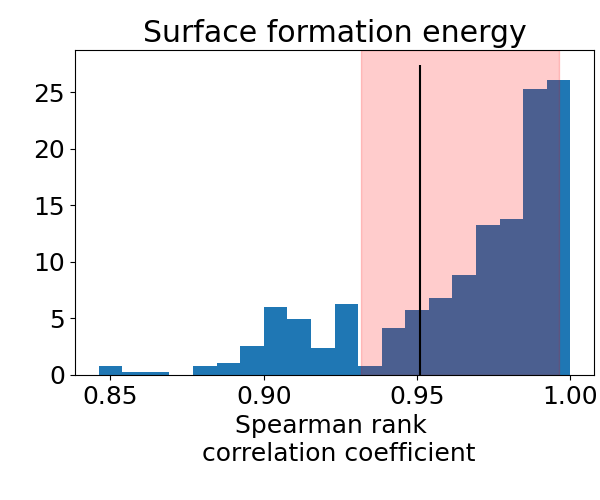}
\caption{\label{fig:correlation} Spearman rank correlation analysis for various quantities. Blue: histogram of rank correlations between MLE and ensemble models; Black line: rank correlation between DFT and MLE. The shaded area corresponds to a one standard deviation interval around the mean of the UQ ensemble.}
\end{figure}

Furthermore, in addition to information regarding the absolute accuracy of the predictions, it is often desirable to establish whether the MLIAPs can be expected to predict the relative ordering of certain properties across different phases, e.g., of the formation energy which determines the most thermodynamically stable phase at low temperature. Fig.\ \ref{fig:correlation} a) and b) demonstrate that the distribution of Spearman rank correlation coefficients between MLE and members of the UQ ensemble (blue histograms) provides representative estimates of the actual correlation between MLE and the reference data (black vertical line): while most potentials agree with the MLE with regards to the ordering of the formation energies, the relative ordering of the equilibrium volumes shows a much broader distribution. In both cases, the Spearman correlation coefficient between MLE and the reference is contained within a one standard deviation interval around the ensemble to MLE mean. This is a very desirable feature, as it enables the end-user to establish confidence on the accuracy of ranked comparisons without access to reference data.

\begin{figure}[!htb]
\includegraphics[width=0.49\columnwidth]{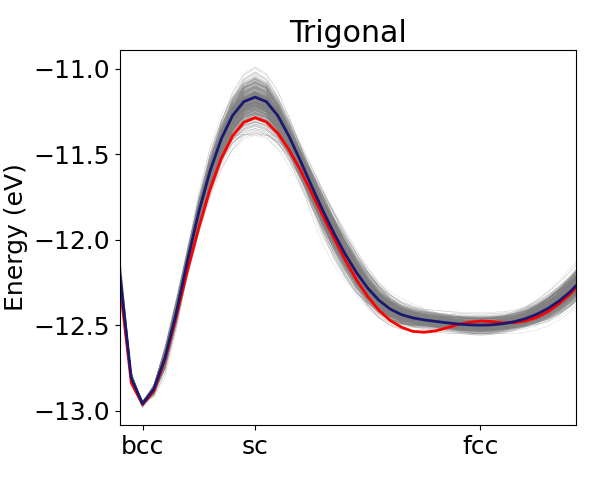}
\includegraphics[width=0.49\columnwidth]{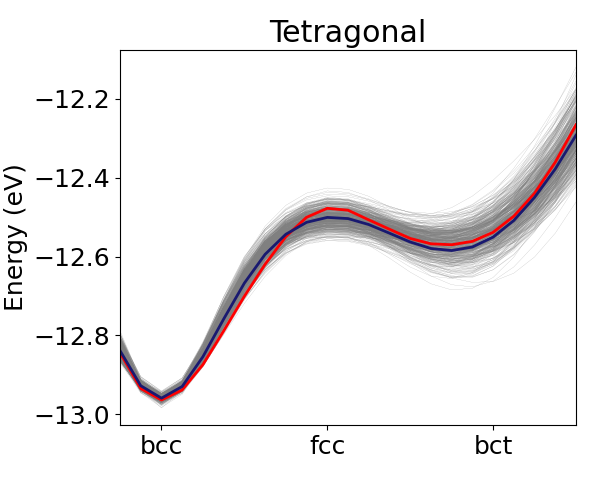}
\includegraphics[width=0.49\columnwidth]{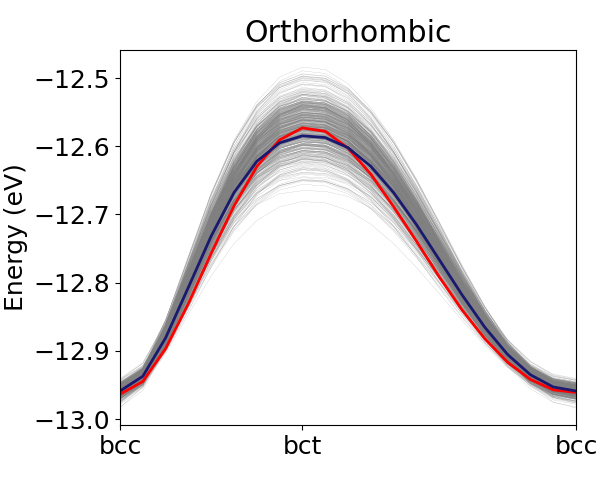}
\includegraphics[width=0.49\columnwidth]{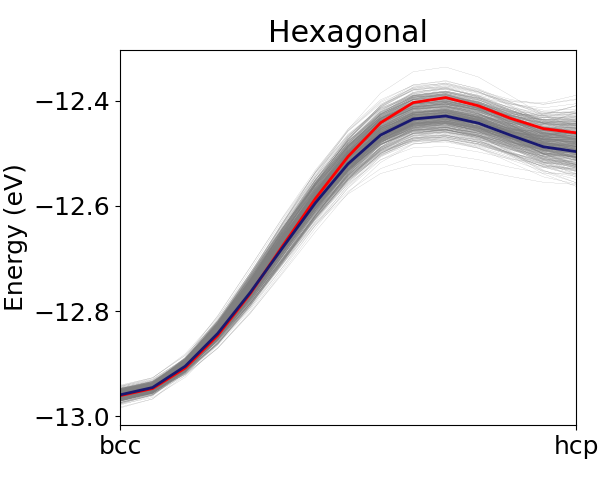}

\caption{\label{fig:transformation} Transformation paths between different crystal phases. MLE predictions are shown in blue, DFT reference values in red, and ensemble predictions in grey. }
\end{figure}

Transformation pathways between crystal structures are also relevant to the analysis of phase transitions. A range of such continuous transformation paths are reported in Fig.\ \ref{fig:transformation}. The MLE MLIAP closely reproduces reference DFT results for the four paths that were considered. In all cases, the distribution of predictions from the UQ ensemble are tightly concentrated, except for the orthorhombic bcc $\rightarrow$ bct $\rightarrow$ bcc path were the prediction in the bct region are a somewhat broader. In all cases, the UQ ensemble bounds the reference DFT values while providing a representative quantification of the actual error incured by the MLE MLIAP.

\subsection{\label{sec:phonons} Phonons}

Another key indicator of the thermodynamics and dynamics of crystal structures is provided by phonon dispersion relations, which are often prized as they can be correlated with scattering or spectroscopic experiments, as well as quantify contribution of vibration-entropic effects to the thermodynamic stability of different crystal structures.
Note that phonon properties derive from the diagonalization of energy Hessians or dynamical matrices and are therefore determined by second-order derivatives of the energy, which were not explicitly present in the training set.
% Anything interesting to say about the fact that we seem to bound high order derivatives without explicitly having POPS for these? It doesn't seem obvious that we necessarily have to. Maybe use the toy example from the previous paper to explore this? It seems to me that the poly+sine example suggests that we wouldn't bound the gradients nor the curvatures if we use energy POPS alone....

% You: I agree we can’t claim to bound higher derivatives in general but perhaps for phonons the case is a bit different? I.e. we have a strong weighting for bcc stability thus +ve curvatures. For the poly+sine we for sure can’t bound curvatures, as in that case the misspecification is so strong the model cannot even span the required second derivatives. For the MLIAP/phonons, we are less misspecified and fitting on force data. In this case we should be able to do better…. perhaps I can derive something about this. Regarding the error predictions, for me there is a strong role of weighting in the fit, which defines the loss function and thus the POPS ensemble (even though it doesn’t affect the POPS themselves of course). i.e. if we strongly weighted low temperature equilibrium AIMD runs (or something similar) we would have a better MLE for phonons and the cube would span DFT more centrally. It could be interesting to look at the  max/min bounds on the cube (not just samples) also, and see how these change with weighting 

Therefore, POPS ensembles were not explicitly constructed to match elements of the Hessian. Comparison of DFT and MLE results show that the MLIAP performs well at low frequencies, but significantly overestimates the vibrational density of states at high frequencies (c.f., right panel), potentially reflecting the absence of Hessian training data. 
% perhaps we can say “low weighting on vibrational dynamics” to make the point that if data is weighted ow then it may be near the edge of the cube, i.e. in a low density prediction region. 
Correspondingly, the range of spectra predicted by the ensemble is also very broad, suggesting low confidence in the predictions. The UQ ensemble however still correctly bounds the reference spectrum across the whole range of wavevectors.

\begin{figure}[htb!]
\includegraphics[width=0.98\columnwidth]{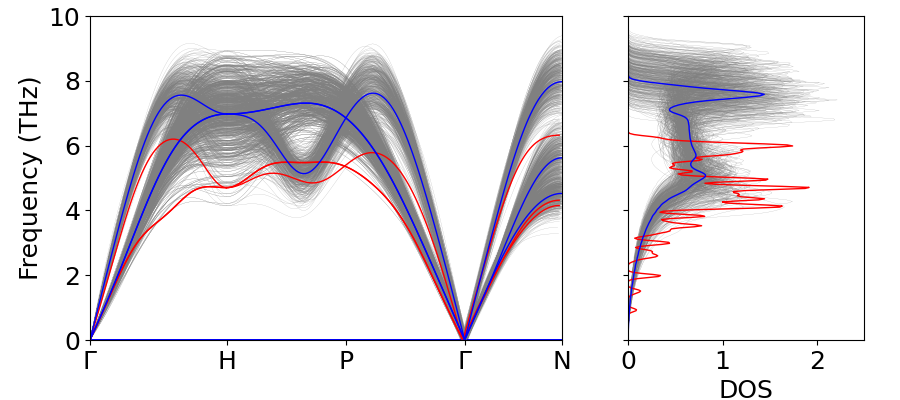}
\caption{\label{fig:cold} Left: phonon dispersion in the BCC phase. Right: vibrational density of states. MLE predictions are shown in blue, DFT reference values in red, and ensemble predictions in grey.}
\end{figure}
%\begin{figure}[!htb]
%\includegraphics[width=0.49\columnwidth]{figures/bcc-cold.png}
%\includegraphics[width=0.49\columnwidth]{figures/a15-cold.png}
%\includegraphics[width=0.49\columnwidth]{figures/c15-cold.png}
%\includegraphics[width=0.49\columnwidth]{figures/c36-cold.png}
%\caption{\label{fig:cold} Cold curves for different crystal structures. MLE predictions %are shown in blue, DFT reference values in red, and ensemble predictions in grey.}
%\end{figure}

\subsection{\label{sec:defects} Defect energetics}

\begin{figure}[!htb]
\includegraphics[width=0.98\columnwidth]{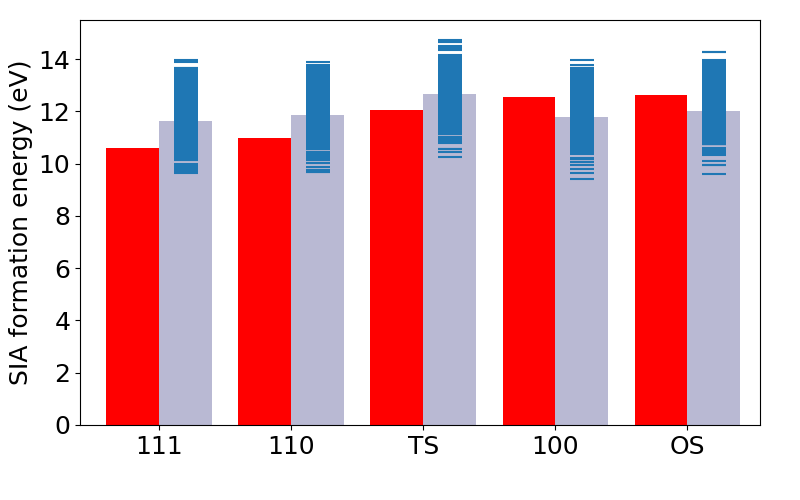}
\caption{\label{fig:sia} Self-interstitial formation energies in the BCC phase.  MLE predictions are shown by blue bars, DFT reference values by red bars, and ensemble predictions by blue lines. }
\end{figure}

Finally, due to their critical role in determining the mechanical properties of engineering materials, the energetics of defects are often key quantities used to train and validate potentials. We considered two classes of defects: self-interstitial atoms (SIAs) --- which are particularly important to understand the behavior of materials under irradiation --- and free surfaces. In both cases, formation energies were obtained self-consistently using the energy-minimizing structures predicted by each potential. The results are presented in Fig.\ \ref{fig:sia} and Table\ \ref{tab:sia}. The energy scale for SIA formation is accurately captured by the MLE model and the ensemble results bound the actual formation energies. The standard deviation of the UQ ensemble provides an excellent statistical representation of the actual error incurred by the MLE. In this case, the 
the formation energies for 110 and OS variants are underestimated by the MLE, leading to a different predicted ordering of the relative defect stabilities. As shown in Fig.\ \ref{fig:correlation}, the distribution of Spearman rank correlation coefficients between MLE and members of the UQ ensemble is also very broad, consistent with the observed ranking disagreement between MLE and reference values; the rank correlation coefficient between the reference and the MLE is found within one standard deviation of the mean of the correlation coefficients between MLE and ensemble.

\begin{table}
    \centering
    \caption{UQ statistics for the formation energetics of SIA in a BCC crystal (c.f., Fig.\ \ref{fig:sia}). MLE error corresponds to the errors of the MLE relative to the DFT reference, Std is the standard deviation of the predictions of $\pi^*_\mathcal{H}$, Low Error is the difference between the smallest prediction in $\pi^*_\mathcal{H}$ and the DFT reference, and High Error is the difference between the largest prediction in $\pi^*_\mathcal{H}$ and the DFT reference. A negative Low Error and a positive High Error indicate that the predictions from the UQ ensemble bracket the reference value.}
    \label{tab:sia}
    \begin{tabular}{ccccc}
        \toprule
        Structure & MLE Error & Std & Low Error & High Error \\
        \midrule
        111 & $1.03\times 10^{ 0 }$ & $7.54\times 10^{ -1 }$ & $-9.70\times 10^{ -1 }$ & $3.33\times 10^{ 0 }$ \\
        110 & $9.09\times 10^{ -1 }$ & $7.58\times 10^{ -1 }$ & $-1.32\times 10^{ 0 }$ & $2.98\times 10^{ 0 }$ \\
        TS & $6.25\times 10^{ -1 }$ & $8.52\times 10^{ -1 }$ & $-2.40\times 10^{ 0 }$ & $2.70\times 10^{ 0 }$ \\
        100 & $-7.57\times 10^{ -1 }$ & $8.30\times 10^{ -1 }$ & $-3.15\times 10^{ 0 }$ & $2.20\times 10^{ 0 }$ \\
        OS & $-6.09\times 10^{ -1 }$ & $8.17\times 10^{ -1 }$ & $-3.23\times 10^{ 0 }$ & $2.12\times 10^{ 0 }$ \\
        \midrule
        Mean $\frac{\mathrm{Error}}{\mathrm{Std}}$ & 0.99 & \multicolumn{2}{c}{Mean $\frac{\mathrm{Error}}{\mathrm{High-Low}}$} & 0.16 \\
        \bottomrule
    \end{tabular}
\end{table}

\begin{figure}[!htb]
\includegraphics[width=0.98\columnwidth]{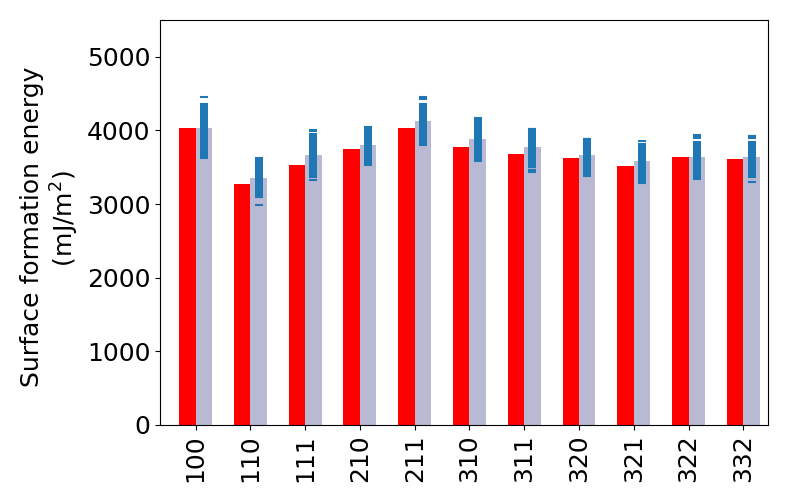}
\caption{\label{fig:surfaces} Surface formation energies in the BCC phase. MLE predictions are shown by blue bars, DFT reference values by red bars, and ensemble predictions by blue lines. }
\end{figure}

\begin{table}
    \centering
    \caption{UQ statistics for the formation energetics of surfaces in a BCC crystal (c.f., Fig.\ \ref{fig:surfaces}). MLE error corresponds to the errors of the MLE relative to the DFT reference, Std is the standard deviation of the predictions of $\pi^*_\mathcal{H}$, Low Error is the difference between the smallest prediction in $\pi^*_\mathcal{H}$ and the DFT reference, and High Error is the difference between the largest prediction in $\pi^*_\mathcal{H}$ and the DFT reference. A negative Low Error and a positive High Error indicate that the predictions from the UQ ensemble bracket the reference value.}
    \label{tab:surfaces}
    \begin{tabular}{ccccc}
        \toprule
        Structure & MLE Error & Std & Low Error & High Error \\
        \midrule
        100 & $-1.19\times 10^{ 0 }$ & $1.32\times 10^{ 2 }$ & $-4.06\times 10^{ 2 }$ & $4.15\times 10^{ 2 }$ \\
        110 & $8.88\times 10^{ 1 }$ & $3.61\times 10^{ 2 }$ & $-2.72\times 10^{ 2 }$ & $1.19\times 10^{ 3 }$ \\
        111 & $1.35\times 10^{ 2 }$ & $3.02\times 10^{ 2 }$ & $-5.41\times 10^{ 2 }$ & $9.17\times 10^{ 2 }$ \\
        210 & $5.72\times 10^{ 1 }$ & $2.71\times 10^{ 2 }$ & $-7.48\times 10^{ 2 }$ & $7.09\times 10^{ 2 }$ \\
        211 & $9.14\times 10^{ 1 }$ & $2.95\times 10^{ 2 }$ & $-1.04\times 10^{ 3 }$ & $4.14\times 10^{ 2 }$ \\
        310 & $1.19\times 10^{ 2 }$ & $2.74\times 10^{ 2 }$ & $-7.76\times 10^{ 2 }$ & $6.82\times 10^{ 2 }$ \\
        311 & $8.27\times 10^{ 1 }$ & $2.57\times 10^{ 2 }$ & $-6.92\times 10^{ 2 }$ & $7.65\times 10^{ 2 }$ \\
        320 & $4.45\times 10^{ 1 }$ & $2.47\times 10^{ 2 }$ & $-6.26\times 10^{ 2 }$ & $8.31\times 10^{ 2 }$ \\
        321 & $7.56\times 10^{ 1 }$ & $2.43\times 10^{ 2 }$ & $-5.21\times 10^{ 2 }$ & $9.37\times 10^{ 2 }$ \\
        322 & $9.52\times 10^{ 0 }$ & $2.37\times 10^{ 2 }$ & $-6.39\times 10^{ 2 }$ & $8.18\times 10^{ 2 }$ \\
        332 & $2.93\times 10^{ 1 }$ & $2.30\times 10^{ 2 }$ & $-6.16\times 10^{ 2 }$ & $8.41\times 10^{ 2 }$ \\
        \midrule
        Mean $\frac{\mathrm{Error}}{\mathrm{Std}}$ & 0.24 & \multicolumn{2}{c}{Mean $\frac{\mathrm{Error}}{\mathrm{High-Low}}$} & 0.045 \\
        \bottomrule
    \end{tabular}
\end{table}

Surfaces are another class of important planar defects that, e.g., control the shape of nanoparticles. Fig.\ \ref{fig:surfaces} and Table\ \ref{tab:surfaces} demonstrates that the MLE MLIAP in provides an adequate representation of the energies of different facets. In this case, the standard deviation of the ensemble prediction conservatively overestimates the actual errors by about a factor of 4, once again providing worst-case bounds that always include the actual reference value.

Fig.\ \ref{fig:correlation} also shows that the ordering of surface energies is robustly captured by the MLE MLIAP, which is corroborated by the narrow distributions of rank correlation coefficients between MLE and members of the UQ ensemble. In these cases also, the distribution of Spearman coefficients is consistent with the very high correlation between the MLE and the reference.

\subsection{\label{sec:barriers} Energy barriers}

\begin{figure}[!htb]
\includegraphics[width=0.98\columnwidth]{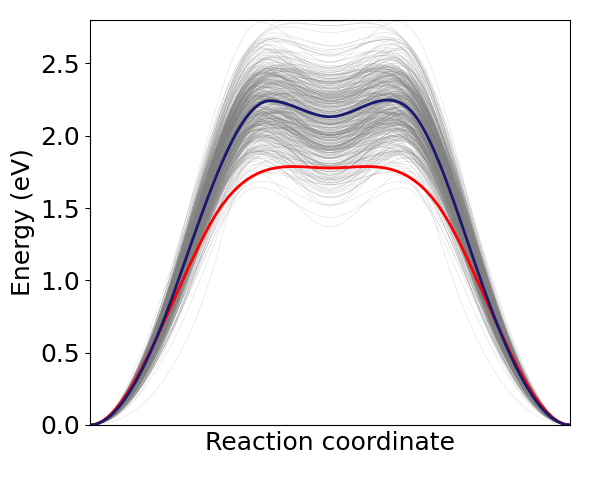}
\caption{\label{fig:vacancy_hop} Minimum energy pathway for a first-neighbor vacancy hop in BCC. MLE predictions are shown in blue, DFT reference values in red, and ensemble predictions in grey. }
\end{figure}

In addition to thermodynamics, an assessment of uncertainty of properties related to defect kinetics is often extremely desirable, especially since kinetic properties can be exponentially sensitive to transition barrier energetics. This makes it extremely important to avoid overly pessimistic UQ, since it can translate into exponentially large differences in predicted characteristic timescales. Furthermore, saddle points are computationally expensive to harvest in large numbers using reference quantum methods, which makes them potentially drastically underrepresented in most training sets for MLIAPs. Fig.\ \ref{fig:vacancy_hop} reports on the performance of the UQ ensemble for the minimum energy pathway of a first neighbor vacancy hop in BCC W. The results show that the MLE overestimates the reference results by a significant margin (about 0.5 eV), but that the UQ ensemble offers a quantitatively appropriate estimation of the error on the energy barrier. Note that the minimum energy pathways were individually reconverged for each MLIAP, and not simply reevaluated along the reference minimum energy pathway.

\subsection{\label{sec:impl}Fast UQ propagation via implicit expansions}
Many material properties, such as defect energetics and energy barriers, 
are calculated via local energy minimization.
In principle, propagation of parameter uncertainty to these properties requires 
brute force repetition of simulations, which quickly becomes 
unfeasible as system size or system count increases. 
In this section, we apply a recently introduced approach to assess the 
predictions from the UQ ensemble by employing the implicit differentiation of 
atomic minima\cite{maliyov2024impl_diff}. The implicit derivative emerges 
by noting a {stationary} atomic configuration $\mathbf{X}^*$ is an \textit{implicit} function of the potential parameters $\mathbf{\Theta}$. As shown in 
Ref.~\cite{maliyov2024impl_diff}, the implicit derivative of atomic 
configurations, $\nabla_{\mathbf{\Theta}} \mathbf{X}^*_{\mathbf{\Theta}}$, can 
be computed efficiently for linear-in-descriptor potentials. This enables the 
calculation of the change in stationary configurations, 
$\Delta \mathbf{X}^*_{\mathbf{\Theta}}$, under relatively small potential 
perturbations, $\Delta\mathbf{\Theta}$, without re-minimization of the system 
for each potential sample. This method is advantageous in scenarios where 
performing molecular statics calculations is expensive due to the large system 
size or a large number of ensemble potentials. 

Here, we apply the implicit approach to two UQ estimation cases presented above: 1) equilibrium volumes of BCC and HCP W phases and 2) minimum energy pathways for a first-neighbor vacancy hop in BCC W. For both cases, implicit derivative of the equilibrium volumes $V^*_{\mathbf{\Theta}}$, $\nabla_{\mathbf{\Theta}} V^*_{\mathbf{\Theta}}$ is sufficient for the predictions. More details of the implicit expansion method and various forms of truncation/approximation are given in Ref.~\cite{maliyov2024impl_diff}.

\begin{figure}[t!]
    \centering
    \includegraphics[width=1\columnwidth]{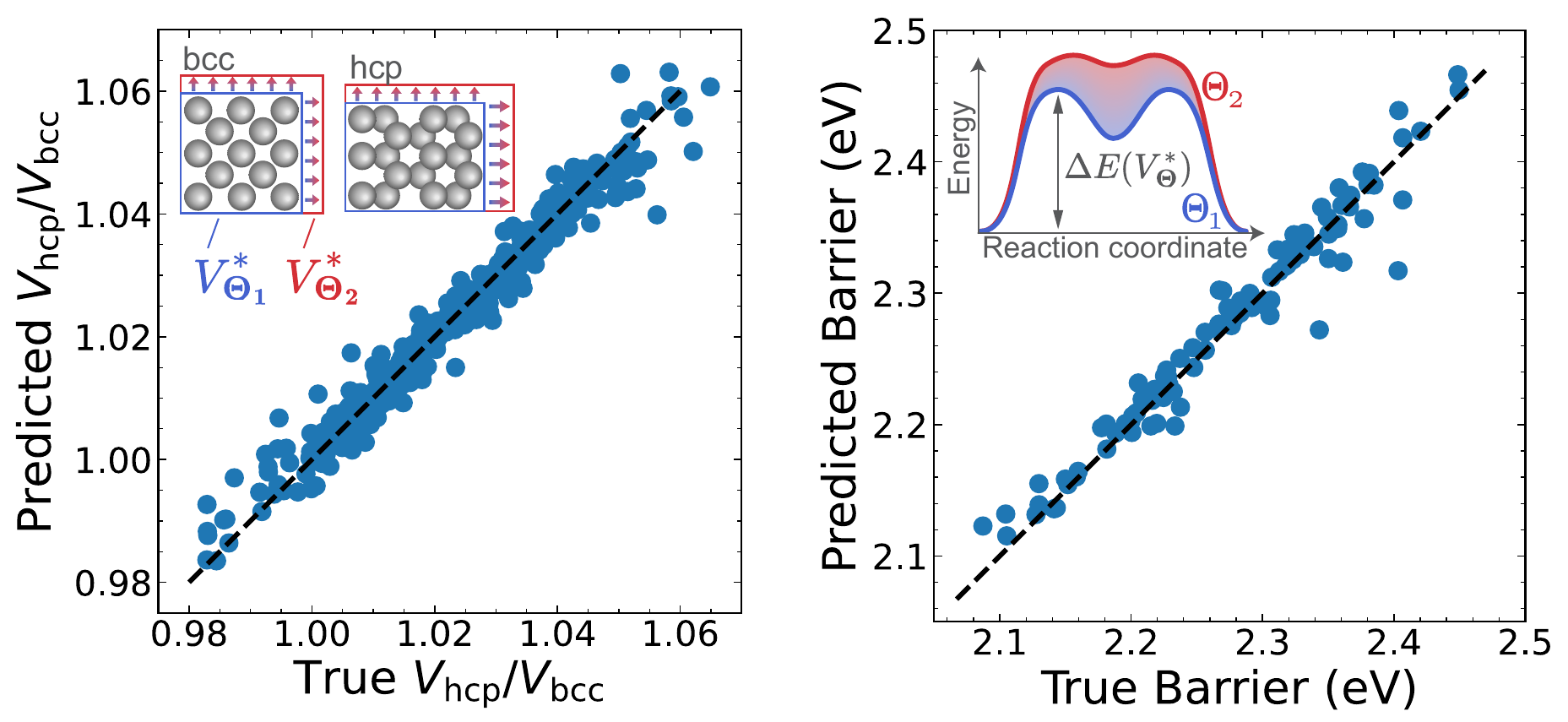}
    \caption{Implicit derivative predictions vs true molecular statics minimization for the UQ ensemble potentials. Left: ratio of equilibrium volumes of HCP and BCC W phases. Right: minimum energy pathway barriers for a first-neighbor vacancy hop in BCC W.
    }
    \label{fig:UQ-impl-diff}
\end{figure}

For UQ of the equilibrium volumes, we first compute the implicit derivatives $\nabla_{\mathbf{\Theta}} V^*_{\mathbf{\Theta}}$ at BCC and HCP minima with the MLE potential. Then, for each potential sample from the UQ ensemble, we predict the BCC and HCP volume change $\Delta V^*_{\mathbf{\Theta}} = \Delta \mathbf{\Theta} \cdot \nabla_{\mathbf{\Theta}} V^*_{\mathbf{\Theta}}$. Left panel of Fig.~\ref{fig:UQ-impl-diff} shows the predicted BCC and HCP volume ratios vs their true values obtained with minimization for potentials from the UQ ensemble. For UQ of the minimum energy pathways, we perform the full calculation with the MLE potential, and identify the initial and saddle-point configurations. We then compute the implicit volume derivative at the \textit{initial} configuration. We predict the energy change at initial and saddle-point configurations using the Taylor expansion for atomic energy with implicit derivative (see Ref.~\cite{maliyov2024impl_diff} for more details). Figure~\ref{fig:UQ-impl-diff}, right panel, shows the implicit derivative predictions of the energy barriers compared to the full pathway calculations. 

For both cases, the implicit derivative technique provides the predictions within less then 4\% of error for both cases. Since the goal of the POPS approach is to provide the worst-case bounds for the quantities of interest, combination of the UQ ensemble potentials with the implicit derivative predictions provides the ultimate efficient scheme for the UQ of the molecular statics properties.
%\section{\label{sec:discussion} Discussion}

\subsection{Application to universal MLIAPs}
Recent message-passing neural network (MPNN) models \cite{batatia2022mace,Bochkarev2024} 
have shown impressive approximation ability to predict atomic energies 
and forces of diverse multi-specie configurations across the periodic table
\cite{batatia2023foundation,deng_2023_chgnet}. There is thus significant interest in assessing 
the accuracy of these `universal' MLIAPs (UMLIAPs), to determine both uncertainty 
in predictions and select optimal training configurations for fine-tuning, 
where additional training data is used to adjust a small subset of model parameters. 

\begin{figure}[!htb]
\includegraphics[width=0.98\columnwidth]{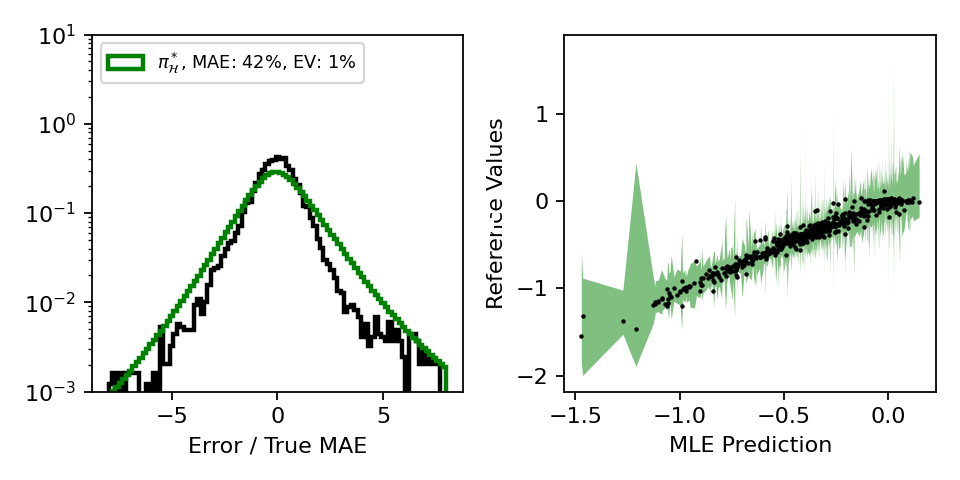}
\caption{\label{fig:mace} 
Characterization of the statistics of pointwise energy errors of the MACE-MPA-0 foundation 
model, obtained from $\pi^*_\mathcal{H}$ applied to the linear corrector (\ref{MACE-correction}). 
Left: distribution of test errors for the MLE against the reference data (black) and 
from $\pi^*_\mathcal{H}$ \textit{ansatz} against the MLE (green). MAE: mean absolute error relative to the minimum loss solution. EV: envelope violation, fraction of points lying outside of max/min bound. Right: parity plot of actual vs MLE predicted energies for a subset of 
points. Shaded areas show min/max range of predictions over all members of $\pi^*_\mathcal{H}$.
}
\end{figure}

The per-atom energy prediction of UMLIAPs $E^i_{MPNN}$ is produced by a readout 
function\cite{batatia2022mace,Bochkarev2024} , which typically 
receives scalar-valued messages from the MPNN. Most fine-tuning schemes 
only adjust parameters in this readout layer for computational efficiency. 
In the framework of this paper, we can therefore treat the scalar-valued 
input to the readout layer as per-atom descriptors ${\bf D}^i$, as in the 
MACE MPNN model\cite{batatia2022mace}. To motivate forthcoming studies 
of misspecification-aware UQ and fine-tuning for UMLIAPs, we applied 
the POPS UQ scheme to a linear corrector for the MACE-MPA-0 foundation 
model\cite{batatia2023foundation}, trained on the \texttt{mptraj}\cite{deng_2023_chgnet} and \texttt{sAlex}\cite{schmidt2023machine} datasets. 
Specifically, we consider a simple linear model in addition to the MACE-MPA-0 prediction, giving a loss function
\begin{equation}
    L(\boldsymbol{\Theta}) = \frac{1}{2}\sum_i\|E^i_{\rm DFT}-E^i_{\rm MACE-MPA-0} - \boldsymbol{\Theta}\cdot{\bf D}_i\|^2,
    \label{MACE-correction}
\end{equation}
where ${\bf D}_i\in\mathbb{R}^{256}$ is the MACE per-atom descriptor vector.
We applied the POPS scheme to obtain a posterior distribution $\pi^*_\mathcal{H}(\boldsymbol{\Theta})$
trained over energies from the \texttt{mptraj} dataset, including
all 89 elements of\texttt{mptraj} with a 50:50 train:test split. 
As shown in figure \ref{fig:mace}, the ability of 
POPS to accurately predict test error distributions and bound 
worst case errors seen for linear MLIAPs is maintained in application to 
linear correctors for UMLIAPs. We observe excellent coverage 
of the test error distribution over at least $\pm4$ standard deviations, 
with a small envelope violation of 1\%. These preliminary results 
show that the general misspecification-aware framework introduced here can 
be applied to recent universal MLIAPs; future work will develop this approach 
both for uncertainty propagation and active learning workflows for UMLIAP fine-tuning.

\section{\label{sec:conclusion} Conclusion}
This paper has investigated uncertainty quantification for the predictions of machine learning interatomic potentials (MLIAPs). We demonstrated application of a novel Bayesian regression approach, POPS\cite{swinburne2025}, that is specifically designed for near-deterministic regression tasks when the aleatoric error is low (e.g., when reference quantum calculations are well converged) and training data is abundant, so that model misspecification errors dominate. 
The effect of this type of error is comparatively under-studied in the Bayesian regression literature, where the focus primarily lies on quantifying the effects of the lack of training data, but is essential to understand uncertainties in conditions typical of the development of modern MLIAPs. The method is extremely computationally efficient for the broad class of MLIAPs that can be expressed as linear combinations of very complex non-linear features, such as the ACE \cite{PhysRevB.99.014104} and SNAP potentials \cite{thompson2015spectral,wood2018extending}, introducing a negligible additional cost to generate a statistically-representative ensemble of MLIAPs. 

The ensemble of potential weights generated by the POPS approach proved extremely adept at quantitatively estimating uncertainties on both pointwise and complex quantities and at bounding worst-case errors. Through an extensive suite of validation tests commonly used to assess MLIAP quality for materials science, including static, dynamic, and kinetic properties of perfect crystals and defects,  we demonstrated that robust uncertainty metrics can be reliably obtained at a low computational cost. This type of approach offers dramatic improvement in the quantitative understanding of uncertainties inherent to atomistic simulations in the ML era. We also demonstrated the POPS 
framework can be applied to bound the error of non-linear models, specifically recent MPNN models\cite{batatia2023foundation}, through 
the use of a linear corrector which will be developed further in future work.
More generally, our study highlights the benefit of principled, misspecification-aware UQ techniques to systematically optimize accuracy/simulation rate tradeoffs, crucial 
to realize the predictive potential of data-driven models.
\section{\label{sec:acknowledgements} Acknowledgments}
We gratefully acknowledge useful discussion with Dr. Peter Hatton and the hospitality of the Institute for Pure and Applied Mathematics (IPAM) at UCLA and of the Institute for Mathematical and Statistical Innovation (IMSI) at the University of Chicago during the conception of this work. DP was supported by the Laboratory Directed Research and Development program of Los Alamos National Laboratory under project number 20220063DR. APAS acknowledges the support from the US Department of Energy through the Exascale Computing Project
(17-SC-20-SC), a collaborative effort of the U.S. Department of Energy Office of Science and the National Nuclear Security Administration and through the G. T. Seaborg Institute under project number 20240478CR-GTS. 
TDS gratefully acknowledges support from ANR grants ANR-19-CE46-0006-1, ANR-23-CE46-0006-1, IDRIS allocation A0120913455, and, with IM, an Emergence@INP grant from the CNRS.
Los Alamos National Laboratory is operated by Triad National Security, LLC, for the National Nuclear Security Administration of U.S. Department of Energy (Contract No. 89233218CNA000001).
\bibliography{main}% Produces the bibliography via BibTeX.

\end{document}